\documentclass[10pt,final,doublecolumn]{IEEEtran}
\usepackage{amsmath}
\usepackage{latexsym}
\usepackage{graphicx}
\usepackage{bbding}
\usepackage{indentfirst}
\usepackage{algorithm,algorithmic}
\usepackage{setspace}
\usepackage{float}
\usepackage{epstopdf}
\usepackage{amssymb}
\usepackage{amsfonts}
\usepackage{enumerate}
\usepackage{multicol}
\usepackage{color}
\usepackage{bm}
\usepackage{amssymb}
\usepackage{stfloats}
\usepackage{epstopdf}
\usepackage{threeparttable}
\usepackage{epstopdf}
\usepackage{threeparttable}
\usepackage{subfigure}
\usepackage{makecell}
\usepackage{cite}
\usepackage{verbatim}
\IEEEoverridecommandlockouts
\allowdisplaybreaks[4]
\begin{document}
	\title{6DMA-Aided Cell-Free Massive MIMO Communication}
\author{Xiaoming Shi, Xiaodan Shao,~\IEEEmembership{Member,~IEEE}, Beixiong Zheng,~\IEEEmembership{Senior Member,~IEEE,} and Rui Zhang,~\IEEEmembership{Fellow,~IEEE}
	\vspace{-26pt}
	
	\thanks{X. Shi is with School of Science and Engineering, The Chinese University of Hong Kong, Shenzhen, Guangdong 518172, China (e-mail: xiaomingshi@link.cuhk.edu.cn).}
	\thanks{X. Shao is with Department of Electrical and Computer Engineering, University of Waterloo, Waterloo, ON N2L 3G1, Canada (e-mail: x6shao@uwaterloo.ca).}
	\thanks{B. Zheng is with School of Microelectronics, South China University of
		Technology, Guangzhou 511442, China (e-mail: bxzheng@scut.edu.cn).}
	\thanks{R. Zhang is with School of Science and Engineering, Shenzhen Research
		Institute of Big Data, The Chinese University of Hong Kong, Shenzhen,
		Guangdong 518172, China. He is also with the Department of Electrical and
		Computer Engineering, National University of Singapore, Singapore 117583
		(e-mail: elezhang@nus.edu.sg). 
		\emph{(Corresponding author: Xiaodan Shao and Rui Zhang.)}}}
	\maketitle
	\begin{abstract}
		In this letter, we propose a six-dimensional movable antenna (6DMA)-aided  cell-free massive multiple-input multiple-output (MIMO) system to fully exploit its macro spatial diversity, where a set of distributed access points (APs), each equipped with multiple 6DMA surfaces, cooperatively serve all users in a given area.
		Connected to a central processing unit (CPU) via fronthaul links, 6DMA-APs can optimize their combining vectors for decoding the users' information based on either local channel state information (CSI) or global CSI shared among them.
		We aim to maximize the average achievable sum-rate via jointly optimizing the rotation angles of all 6DMA surfaces at all APs, based on the users' spatial distribution.
		Since the formulated problem is non-convex and highly non-linear, we propose a Bayesian optimization-based algorithm to solve it efficiently. 
		Simulation results show that, by enhancing signal power and mitigating interference through reduced channel cross-correlation among users, 6DMA-APs with optimized rotations can significantly improve the average sum-rate, as compared to the conventional cell-free network with fixed-position antennas and that with only a single centralized AP with optimally rotated 6DMAs, especially when the user distribution is more spatially diverse.
	\end{abstract}
		\vspace{-4pt}
	\vspace{-10pt}
	\section{Introduction}
	\vspace{-2pt}
	The deployment of dense base stations (BSs) or access points (APs) is the current trend to achieve massive connectivity and seamless coverage for the forthcoming sixth-generation (6G) wireless sensing \cite{sensing} and communication networks \cite{7476821}.
	A promising approach to achieve this goal is the cell-free massive multiple-input multiple-output (mMIMO) system \cite{9586055,chen2022survey}, which consists of a large number of distributed APs connected to one or more central processing units (CPUs) to cooperatively serve all users in a given area without cell boundaries via joint signal encoding/decoding, based on the users' channel state information (CSI) locally estimated at individual APs or globally shared with the CPU.
	However, the existing studies on cell-free mMIMO assume fixed-position antennas (FPAs) at each AP and thus cannot flexibly allocate antenna resources in the network according to the users' spatial distribution, which inevitably limits the achievable rates of cell-free mMIMO systems.
	
	To overcome this limitation, the recently proposed six-dimensional movable antenna (6DMA) technique provides a promising solution \cite{shao20246d,6dma_dis,shaoaga}.
	In 6DMA systems, distributed antennas/arrays can move and rotate individually in a given three-dimensional (3D) space, cooperatively varying their steering vectors for achieving favorable channel conditions.
	Besides, 6DMA-empowered BS/AP can adaptively allocate its antenna resources based on the long-term/statistical user channel distribution, thus fully leveraging the antennas' directionality, array gain, and spatial multiplexing gain to significantly improve network capacity.
	6DMA is thus more advantageous than not only traditional FPAs, but also existing  fluid/movable antennas \cite{9650760,10286328,10318061}, which can adjust antenna positions on a finite line/surface only to mitigate the small-scale channel fading based on the instantaneous CSI. 
	By applying 6DMA to cell-free mMIMO, the 6DMA surfaces at all APs can jointly optimize their positions and rotations, further exploiting the network’s macro spatial diversity to enhance its capacity. 
	This motivates our current work as 6DMA-aided cell-free mMIMO has not been rigorously investigated in the literature yet.

	Specifically, this letter studies a 6DMA-aided cell-free mMIMO system in the uplink transmission, where multiple 6DMA-APs jointly decode signals from all distributed users in a given area. 
	For ease of implementation, we consider a circular 6DMA architecture where all 6DMA surfaces at each AP can freely move along a circular track \cite{ming}. 
	Each AP applies the minimum-mean-square-error (MMSE)-based combining vectors over its antennas according to either global or local CSI to maximize the signal-to-interference-and-noise ratio (SINR) at the CPU  for decoding the information of each user. 
	To avoid frequent movement of 6DMA surfaces, we aim to maximize  the average achievable sum-rate of all users by jointly optimizing the rotation angles of all 6DMA surfaces at all APs, based on {\it a priori} known users' spatial distribution.
	We propose an efficient Bayesian optimization (BO)-based algorithm to solve the formulated non-convex optimization problem.
	Numerical results show that the proposed 6DMA-aided cell-free network outperforms both the FPA-based cell-free network and the 6DMA-aided single AP given the same total number of antennas in terms of average sum-rate.
	This performance gain is achieved by optimally allocating 6DMA surfaces at all APs to match the (non-uniform) users' spatial distribution, which enhances the desired signal power while suppressing the undesired interference more effectively.
	
	\textit{Notations}: $(\cdot)^H$ and $(\cdot)^T$ denote conjugate transpose and transpose, respectively, ${\bf I}_{N}$ denotes the $N \times N$ identity matrix, $\max\{ \bf x\}$ and $\min\{ \bf x\}$ denote the maximum and minimum elements of vector $\bf x$, respectively, $[{\bf x}]_k$ denotes the $k$th element of vector $\bf x$, $\otimes$ denotes the Kronecker product, ${\bf 0}_t $ denotes the $t \times 1$ all-zero vector, and  ${\bf{1}}(\cdot)$ denotes the indicator function, defined as ${\bf 1}(P)=1$ if $P$ is true and $0$ otherwise.

\vspace{-3 mm}
\section{System Model}
\vspace{-1 mm}
\begin{figure*}[t]
	\centering
	\begin{minipage}[t]{0.5\textwidth}
		\centering
		\includegraphics[width=2.5in]{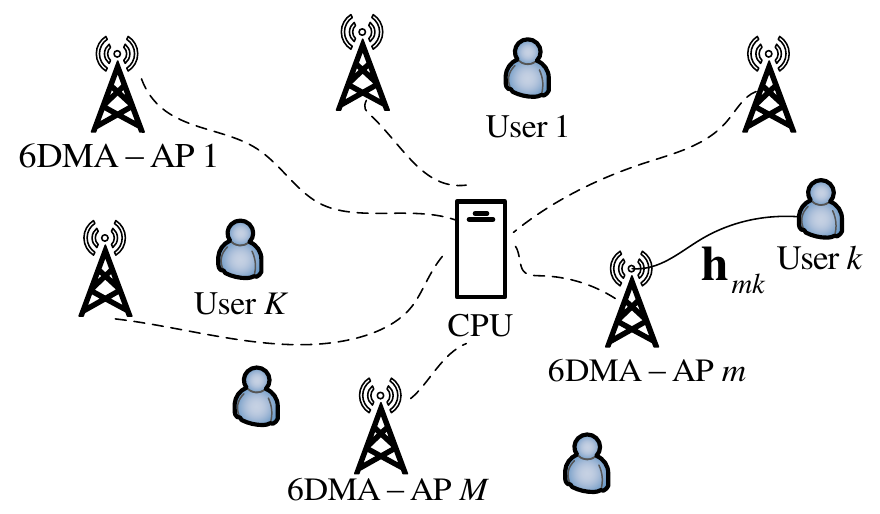}
	\end{minipage}%
	\begin{minipage}[t]{0.5\textwidth}
		\centering
		\includegraphics[width=2.6 in]{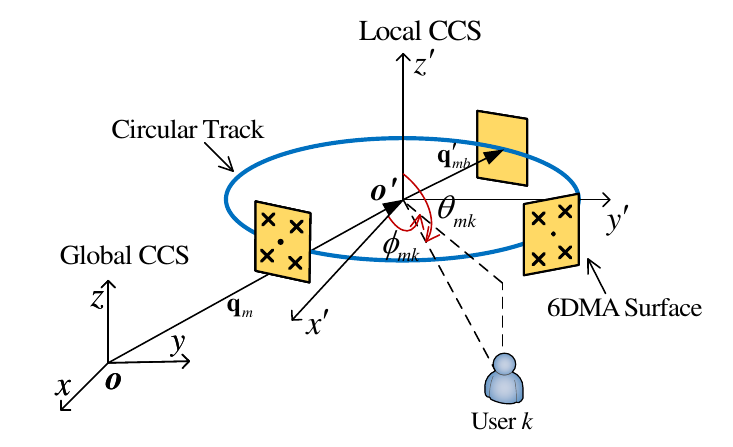}
	\end{minipage}\\[-15pt]
	\begin{minipage}[t]{0.45\textwidth}
		\caption{Illustration of a 6DMA-aided cell-free network with a set of distributed 6DMA-APs connected to the CPU.} \label{fig1}
	\end{minipage}%
		\hspace{5 mm}
	\begin{minipage}[t]{0.45\textwidth}
		\caption{Geometry of the 6DMA-AP.} \label{fig2}
	\end{minipage}
	\vspace{-0cm}
\end{figure*}

As shown in Fig. \ref{fig1}, we consider a cell-free network consisting of $K$ users, denoted by the set $\mathcal{K}=\{1,2,\cdots,K\}$, each equipped with a single isotropic antenna, and $M$ 6DMA-APs, denoted by the set $\mathcal{M}=\{1,2,\cdots,M\}$.
All 6DMA-APs are connected via fronthaul links to a CPU, which is responsible for the multi-AP joint signal processing.
\vspace{-4 mm}
\subsection{6DMA-AP Architecture}
\vspace{-1 mm}
Each 6DMA-AP consists of $B$ 6DMA surfaces, denoted by the set $\mathcal{B} \!=\!\{1,2,\cdots,B\}$.
Each 6DMA surface is assumed to be a uniform planar array (UPA) with a given size, consisting of $N \!=\! N_h \!\times\! N_v$ antennas, where $N_h$ and $N_v$ denote the number of antennas in the horizontal and vertical directions, respectively.
For each 6DMA-AP, all $B$ 6DMA surfaces are connected to a local processor at this AP via flexible cables, and can move along a circular track that is parallel to the ground with a radius of $r$ (see Fig. \ref{fig2}).{\footnote{Although jointly positioning and rotating 6DMA surfaces offers the greatest flexibility, its implementation may be challenging in practice due to the drastic changes required for existing APs/BSs in wireless networks, which may lead to significantly increased infrastructure cost.}}
For simplicity, we set the circular track's center as the reference position of 6DMA-AP $m$, $m\!\in\!\mathcal{M}$, and denote its position in the global Cartesian coordinate system (CCS) $o-xyz$ as ${\bf q}_m\!=\![x_m,y_m,h]^T$.
For each 6DMA-AP, we define a local CCS, denoted by $o'-x'y'z'$, with ${\bf q}_m$, $m\in\mathcal{M}$, as the origin $o'$, and the axes of the local CCS are oriented the same as those of the global CCS.
Furthermore, we define the rotation angle of 6DMA surface $b$ at AP $m$, denoted by $\varphi_{mb}\in [0,2\pi)$, $m\in\mathcal{M}, b\in\mathcal{B}$, as the azimuth angle measured from the positive $x'$-axis in the local CCS of AP $m$ to the center of 6DMA surface $b$.
Thus, the position of the center of 6DMA surface $b$ at AP $m$ in its local CCS can be expressed as 
${{\bf q'}_{mb}}({\varphi _{mb}}) = {[r\cos ({\varphi _{mb}}),r\sin ({\varphi _{mb}}),0]^T}$.
\vspace{-4 mm}
\subsection{Channel Model}
\vspace{-1 mm}
We denote $\phi_{mk}\in \left[0,2\pi\right)$ and $\theta_{mk}\in \left(\frac{\pi}{2},\pi\right)$ as the azimuth and elevation angles, respectively, of the signal from user $k$ arriving at the reference position of AP $m$ (see Fig. \ref{fig2}).
Specifically, for user $k$ located at ${\bf u}_k \! =\!{\left[ {{\alpha _k},{\beta _k},0} \right]^T}$ in the global CCS, $\phi_{mk}$ and $\theta_{mk}$ are given by
${\phi _{mk}} \! = \! {\arctan \big( {\frac{{{{{\beta _k} - {y_m}}}}}{{{{\alpha _k} - {x_m}}}}} \big)} + \pi\cdot{\bf 1}({\alpha _k} \! < \! {x_m}) + 2 \pi\cdot{\bf 1}({\alpha _k} \!>\!  {x_m}){\bf 1}({\beta _k} \!>\!  {y_m})$
and ${\theta _{mk}} = \arctan ( {{h}/{{\sqrt {{( {{\alpha _k} - {x_m}} )^2} - {( {{\beta _k} - {y_m}} )^2}} }}} ) + \frac{\pi }{2}$.
We assume that the inter-antenna distance on each 6DMA surface is $\lambda/2$, where $\lambda$ is the carrier wavelength.
Given $\phi_{mk}$ and $\theta_{mk}$, the array response vector ${\bf f}_{mbk}\in\mathbb{C}^{N\times1}$ of 6DMA surface $b$ at AP $m$ is given by \cite{ming}
\vspace{-1.5 mm}
\begin{equation}\vspace{-1.5 mm}
	{\bf f}_{mbk}(\varphi_{mb})=\sqrt{g(\varphi_{mb})}{e^{j\rho_{mbk}(\varphi_{mb})  }}{\bf a}_{h}({\varphi}_{mb}) \otimes {\bf a}_{v},
\end{equation}
where $[{\bf a}_{h}({\varphi}_{mb})]_n \!=\!{{{e^{j\pi ( {\frac{{{N_h} \!+ 1}}{2} - n} )\sin ( {{\varphi_{mb}} - {\phi _{mk}}} )\sin ( {{\theta _{mk}}} )}}}}$, $[{\bf a}_{v}]_n \!=\!{{e^{j\pi ( {\frac{{{N_v} + 1}}{2} - n} )\cos ( {{\theta _{mk}}} )}}}$,
${\bf a}_{h}({\varphi}_{mb})\!\in\!\mathbb{C}^{N_h\times1}$, ${\bf a}_{v}\!\in\!\mathbb{C}^{N_v\times1}$, $\rho_{mbk}(\varphi_{mb})=\frac{{2\pi r }}{{{\lambda }}}\cos(\varphi_{mb}-\phi_{mk})\sin(\theta_{mk})$ denotes the phase difference between the center of 6DMA $b$ and the reference position of AP $m$, and $g(\varphi_{mb})$ denotes the effective antenna gain for 6DMA surface $b$ at AP $m$.
For the purpose of exposition, we assume the line-of-sight (LoS) channel between each user and each AP, while the results of this letter can be extended to the general multi-path channel model \cite{6dma_dis}.  
Under this assumption, the channel vector ${\bf h}_{mk}\!\in\!\mathbb{C}^{NB\times1}$ from user $k$ to AP $m$ is given by
\vspace{-2 mm}
\begin{equation}\label{channel model}\vspace{-2 mm}
	{\bf h}_{mk}({  {\boldsymbol{ \varphi} }_m})\!=\!\sqrt {{\nu _{mk}}} e^{-j\frac{2\pi d_{mk}}{\lambda}}\! {\left[ {{\bf f}_{m1k}^T(\varphi_{m1}), \cdots\! ,{\bf f}_{mBk}^T(\varphi_{mB})} \right]^T},
\end{equation}
where $\nu_{mk}={\nu_0}/{d_{mk}^2}$ denotes the large-scale channel power gain between AP $m$ and user $k$, with $\nu_0$ representing the power gain at the reference distance of $1$ meter (m), and ${d_{mk}}$ representing the distance from user $k$ to the reference position of AP $m$.
The rotation angle vector ${\boldsymbol{ \varphi} }_m=\left[\varphi_{m1},\cdots,\varphi_{mB}\right]^T$ is arranged in ascending order, i.e., $\varphi_{m1}<\varphi_{m2}<\cdots<\varphi_{mB}$.

\vspace{-4 mm}
\subsection{Signal Model}
\vspace{-1 mm}
Let $s_k$ denote the transmit signal of user $k$ with unit power.
The received uplink signal ${{\bf y}_m}\!\in\! \mathbb{C}^{NB\times1}$ at AP $m$ is given by \vspace{-2 mm}
\begin{equation} \label{received signal}\vspace{-2 mm}
	{{{\bf y}_m}} = \sum\limits_{k = 1}^K {{{\bf h}_{mk}}\left( {  {\boldsymbol{ \varphi} }_m} \right)\sqrt{p_0}{s_k} + {{\bf n}_m}} ,
\end{equation}
where $p_0$ denotes the transmit power, and ${\bf n}_m\in \mathbb{C}^{NB\times1}$ denotes the independent and identically distributed (i.i.d.) complex additive white Gaussian noise (AWGN) vector with zero mean and average power $\sigma ^2$.
Each AP applies a receiver combining vector, denoted by ${\bf v}_{mk}\in\mathbb{C}^{NB\times1}$, for user $k$ to obtain ${\bf v}_{mk}^H{\bf y}_m$, which is sent back to the CPU.
The CPU then sums up  ${\bf v}_{mk}^H{\bf y}_m$'s from all APs to obtain the signal for decoding the information of user $k$, which is given by  \vspace{-1.5 mm}
\begin{equation}\vspace{-1.5 mm}
	\begin{aligned}
		{r_k} &= \sum\limits_{m = 1}^M {\bf v}_{mk}^H{{\bf y}_m}  = \bigg( {\sum\limits_{m = 1}^M {{\bf v}_{mk}^H{{\bf h}_{mk}({\boldsymbol \varphi}_m)}} } \bigg){s_k} \\[-2mm]
		&\ \ \ + \sum\limits_{i = 1,i \ne k}^K {\bigg( {\sum\limits_{m = 1}^M {{\bf v}_{mk}^H{{\bf h}_{mi}({\boldsymbol \varphi}_m)}} } \bigg)} {s_i}  + \sum\limits_{m = 1}^M {{\bf v}_{mk}^H{{\bf n}_m}} \\[-2mm] 
		& = {\bf v}_k^H{{\bf h}_k}({\boldsymbol \varphi}){s_k} + \sum\limits_{i = 1,i \ne k}^K {{\bf v}_k^H{{\bf h}_i}({\boldsymbol \varphi}){s_i}}  + {\bf v}_k^H{\bf n},
	\end{aligned}
\end{equation}
where ${\bf h}_k=\left[{\bf h}_{1k}^T,\cdots,{\bf h}_{Mk}^T  \right]^T\in\mathbb{C}^{NBM\times1}$ denotes the collective channel vector from all APs to user $k$, ${\boldsymbol \varphi}=\left[{\boldsymbol \varphi}_1^T,\cdots,{\boldsymbol \varphi}_M^T\right]^T\in\mathbb{R}^{BM\times1}$ denotes the collective rotation-angle vector, ${\bf v}_k=\left[{\bf v}_{1k}^T,\cdots,{\bf v}_{Mk}^T  \right]^T\in\mathbb{C}^{NBM\times1}$ denotes the collective combining vector, and ${\bf n}=\left[{\bf n}_1^T,\cdots,{\bf n}_M^T \right]^T\in\mathbb{C}^{NBM\times1}$ collects all the noise vectors. 
Thus, the SINR at the CPU for decoding the information of user $k$ is given by\vspace{-1.5 mm}
\begin{equation}\label{SINR}\vspace{-1.5 mm}
{\gamma}_k({\boldsymbol \varphi}) = {\frac{{p_0{{\left| {{\bf v}_k^H{{\bf h}_k}({\boldsymbol \varphi})} \right|}^2}}}{{\sum\limits_{i = 1,i \ne k}^K {}  {p_0{{\left| {{\bf v}_k^H{{\bf h}_i}({\boldsymbol \varphi})} \right|}^2} + {\sigma ^2}{{\left\| {{{\bf v}_k}} \right\|}^2}} }}}.
\end{equation}

Note that in cell-free networks, APs can operate either independently using their respective locally estimated CSI or cooperatively by sharing their CSI with the CPU.
Hence, this letter considers two commonly adopted combining techniques for APs with different levels of cooperation among them, i.e., the localized MMSE (LMMSE) combining and the centralized MMSE (CMMSE) combining \cite{chen2022survey}.
Specifically, in the case of LMMSE combining, each AP estimates the signal of user $k$ using local CSI, and the local estimates from all APs are equally combined at the CPU to decode its information.
Assuming perfect local CSI at all APs, the LMMSE combining vector at AP $m$ for user $k$ is given by
\vspace{-1.5 mm}
\begin{equation}\vspace{-1.5 mm}
	{{\hat {\bf v}}_{mk}}\!\left({\boldsymbol \varphi}_m \right)\! =\!\! {\bigg(p_0 {\sum\limits_{i = 1}^K {{\bf h}_{mi}({\boldsymbol \varphi}){\bf h}_{mi}^H({\boldsymbol \varphi})\! +\! {\sigma ^2}{{\bf I}_{NB}}} }\! \bigg)^{\!\!- 1}}\!\!\!{\bf h}_{mk}({\boldsymbol \varphi}),
\end{equation}
where ${\hat{\bf v}}_k\left({\boldsymbol \varphi} \right)=\left[{\hat{\bf v}}_{1k}^T\left({\boldsymbol \varphi}_1 \right),\cdots,{\hat{\bf v}}_{Mk}^T \left({\boldsymbol \varphi}_M \right) \right]^T$.
In the case of CMMSE combining, all APs forward their locally estimated CSI of all users to the CPU, which designs the combining vectors at all APs for all users to achieve higher SINRs than LMMSE combining. 
Assuming perfect global CSI at the CPU, the CMMSE combining vector for user $k$ is given by
\vspace{-2 mm}
\begin{equation}\vspace{-1 mm}
	{\tilde{\bf v}}_k\left({\boldsymbol \varphi} \right) = {\bigg( {p_0\sum\limits_{i = 1}^K{\bf h}_i({\boldsymbol \varphi}){\bf h}_i^H({\boldsymbol \varphi}) + {\sigma ^2}{{\bf I}_{NBM}}} \bigg)^{ - 1}}{\bf h}_k({\boldsymbol \varphi}).
\end{equation}

\vspace{-3 mm}
\section{Problem Formulation}
\vspace{-1 mm}
Based on (\ref{SINR}), the average sum-rate of all users is given by \vspace{-1.5 mm}
\begin{equation}\label{sum-rate1}\vspace{-1.5 mm}
	R({\boldsymbol \varphi}) = \mathbb{E}_{{\Phi}}\bigg[\sum\limits_{k = 1}^K {{{\log }_2}\left( {1 + {\gamma}_k({\boldsymbol \varphi})} \right)}\bigg] ,
\end{equation}
in bits per second per Hertz (bps/Hz), where the expectation $\mathbb{E}[\cdot]$ is with respect to the point process $\Phi$, which models the users' spatial distribution.
The exact average sum-rate in \eqref{sum-rate1} is hard to obtain, hence we use Monte Carlo method to approximate it \cite{ming}. 
Specifically, we generate $\Upsilon$ independent realizations of the number of users and their positions based on $\Phi$, compute the sum-rate for each realization, and then take the sample mean of the sum-rates obtained.
Thus, the average sum-rate in (\ref{sum-rate1}) can be approximated as \vspace{-1.5 mm}
\begin{equation}\label{sum-rate2}\vspace{-1.5 mm}
		\hat{R} ({\boldsymbol \varphi}) = \dfrac{1}{\Upsilon}\sum\limits_{\upsilon=1}^{\Upsilon}\sum\limits_{k = 1}^{K_\upsilon}{{{\log }_2}\left( {1 + {\gamma}_{k,\upsilon}({\boldsymbol \varphi}) } \right)},
\end{equation}
where ${\gamma}_{k,\upsilon}({\boldsymbol \varphi}) = {\gamma}_{k}({\boldsymbol \varphi})|\{{{{\bf h}_{k,\upsilon}}}, {{{\bf v}_{k,\upsilon}}}\}$ denotes the SINR of user $k$ in the $\upsilon$th realization, given the corresponding number of users $K_\upsilon$, the collective combining vector ${\bf v}_{k,\upsilon}$, and the collective channel vector ${\bf h}_{k,\upsilon}$ of user $k$.

We aim to maximize the approximate average achievable sum-rate of the 6DMA-aided cell-free network by jointly optimizing the rotation angles of all 6DMA surfaces at $M$ APs.
The optimization problem is formulated as \vspace{-1.5 mm}
\begin{align}
	\underset{{  {\boldsymbol{ \varphi} }}}{\text{max}}&~~{\hat{R}}({  {\boldsymbol{ \varphi} }}) \label{Problem}\\
	\text {s.t.}&~ {{\varphi _{m(b+1)}} \!-\! {\varphi _{mb}}}  \geqslant  \delta,\forall m \in \mathcal{M},b \! =\! 1,\!\cdots \!,B \! - \! 1, \tag{\ref{Problem}{a}} \label{Problema}\\
	&~{{\varphi _{m1}} + 2\pi - {\varphi _{mB}}}  \geqslant  \delta,\forall m \in \mathcal{M}, \tag{\ref{Problem}{b}}\label{Problemb}\vspace{-1 mm}
\end{align}
where constraints (\ref{Problema}) and (\ref{Problemb}) avoid overlapping between any two adjacent 6DMA surfaces at the same AP by limiting the minimum rotation angle difference between them, denoted by $\delta$. 
Problem (\ref{Problem}) is a non-convex optimization problem due to its non-concave and highly non-linear objective function (\ref{sum-rate2}) over ${\boldsymbol{ \varphi} }$.
Even if we can use the gradient decent-based method \cite{shao20246d} to solve problem (\ref{Problem}) sub-optimally,
the gradient computation of function (\ref{sum-rate2}) over ${\boldsymbol{ \varphi} }$ is computationally prohibitive.
As such, we propose an alternative BO-based algorithm for solving problem (\ref{Problem}) to avoid the direct gradient computation of its objective function, as elaborated in the next section.

\vspace{-2 mm}
\section{Proposed Algorithm}
\vspace{-0.5 mm}
The proposed BO-based algorithm aims to maximize the objective function  ${\hat{R}}({  {\boldsymbol{ \varphi} }})$ by iteratively refining a Gaussian process (GP)-based surrogate model and selecting new observation points via an expected improvement (EI)-based acquisition function. Each iteration consists of 
i) updating the posterior distribution for any candidate point of the objective using the GP; 
ii) determining the next observation point by maximizing the EI-based acquisition function, which is constructed according to the posterior distribution.
The detailed steps are elaborated in the following.

\vspace{-3 mm}
\subsection{GP-Based Surrogate Model}
\vspace{-1 mm}
We define a GP with respect to the zero-mean Gaussian random variable ${\boldsymbol \varphi}$, characterized by a covariance function $\kappa({\boldsymbol \varphi}_i,{\boldsymbol \varphi}_j)\!=\!\exp(-\frac{1}{2}\left\| {{\boldsymbol \varphi}_i-{\boldsymbol \varphi}_j} \right\|^2)$ to quantify the similarity between any two points ${\boldsymbol \varphi}_i$ and ${\boldsymbol \varphi}_j$.
Given a sample set $\mathcal{D}_S \! \triangleq \!\{ ({\boldsymbol \varphi}^{(i)},{\hat{R}}({\boldsymbol \varphi}^{(i)}))  \}_{i=1}^S$ with $S$ sample points ${\tilde{\boldsymbol \varphi}}\!\triangleq\!\big[{\boldsymbol \varphi}^{(1)},\cdots,{\boldsymbol \varphi}^{(S)}\big]^T$ and their corresponding  observed values ${{\bf r}}({\tilde{\boldsymbol \varphi}}) \!\triangleq\! \big[{\hat{R}}({\boldsymbol \varphi}^{(1)}),\cdots,{\hat{R}}({\boldsymbol \varphi}^{(S)})\big]^T$, the covariance matrix of ${\tilde{\boldsymbol \varphi}}$, denoted by ${\bf K}_S \!\in\!\mathbb{R}^{S \times S}$, is defined by $[{\bf K}_S]_{i,j}\! \triangleq \!\kappa({\boldsymbol \varphi}^{(i)},{\boldsymbol \varphi}^{(j)}),i,j \!\in\! \{1,\cdots,S\}$.
According to the definition of GP, for any next sample point, denoted by ${\boldsymbol{ \varphi} }^{(S+1)}$, its predicted value, denoted by ${\hat{R}}^*({  {\boldsymbol{ \varphi} }^{(S+1)}})$, and
${{\bf r}}({\tilde{\boldsymbol \varphi}})$ are jointly Gaussian distributed, i.e., $[{{\bf r}}({\tilde{\boldsymbol \varphi}}),{\hat{R}}^*({  {\boldsymbol{ \varphi} }^{(S+1)}})]^T \!\sim\! \mathcal{N}({\bf 0}_{S+1},{\bf K}_{S+1})$.
Thus, using the conditional properties of Gaussian distribution, the posterior distribution of ${\hat{R}}^*({  {\boldsymbol{ \varphi} }^{(S+1)}})$ is given by \cite{10234224}\vspace{-1.5 mm}
\begin{equation}\label{predictive distribution}\vspace{-1.5 mm}
	{\hat{R}}^*({  {\boldsymbol{ \varphi} }^{(S+1)}}) | \mathcal{D}_S \sim
	{\mathcal{N}}\big( {{\mu}}({  {\boldsymbol{ \varphi} }^{(S+1)}}),\zeta^2({  {\boldsymbol{ \varphi} }^{(S+1)}}) \big),
\end{equation}
with its mean ${{\mu}}({  {\boldsymbol{ \varphi} }^{(S+1)}}) \!=\! {\bf k}_{S}^T{\bf K}_S^{-1}{{\bf r}}({\tilde{\boldsymbol \varphi}})$ and variance $\zeta^2({  {\boldsymbol{ \varphi} }^{(S+1)}})\! =\! \kappa ({  {\boldsymbol{ \varphi} }^{(S+1)}},{  {\boldsymbol{ \varphi} }^{(S+1)}})\!-\!{\bf k}_{S}^T{\bf K}_S^{-1}{\bf k}_{S}$,
where ${\bf k}_{S} = \big[\kappa({\boldsymbol \varphi}^{(1)}, {\boldsymbol \varphi}^{(S+1)}), \dots, \kappa({\boldsymbol \varphi}^{(S)}, {\boldsymbol \varphi}^{(S+1)})\big]^T$ denotes the covariance vector between ${\boldsymbol \varphi}^{(S+1)}$ and the sample points in ${\tilde{\boldsymbol \varphi}}$.

\vspace{-2.5 mm}
\subsection{EI-Based Acquisition Function}
\vspace{-1 mm}
To select the next sample point ${\boldsymbol{ \varphi} }^{(S+1)}$, we aim to maximize the EI over the best observed value known so far, i.e., maximizing $\mathcal{L}({\boldsymbol{ \varphi} })\! =\! \mathbb{E}[ \max \{0,{\hat{R}}(\boldsymbol{ \varphi}) - {\hat{R}}^+ \} | \mathcal{D}_S]$, where ${\hat{R}}^+ \!=\! {\hat{R}}({\boldsymbol \varphi}^{(i^+)})$ denotes the best observed value in the current sample set $\mathcal{D}_S$ with the index $i^+ \!=\! \arg \max_{i \in \{1, \cdots, S\}} {\hat{R}}({\boldsymbol \varphi}^{(i)})$.
Based on the posterior distribution in (\ref{predictive distribution}),
$\mathcal{L}({\boldsymbol{ \varphi} })$ can be expressed in closed form, given by \cite{9784826}\vspace{-1.5 mm}
\begin{equation}\label{AF}\vspace{-1.5 mm}
	\mathcal{L}({\boldsymbol{ \varphi} }) \!= \!\left\{ {\begin{array}{*{20}{l}}
			{\!\!(\mu({\boldsymbol{ \varphi} })\!-\!{\hat{R}}^+)F(Z)\!+\zeta({\boldsymbol{ \varphi} })f(Z),}&{\!\!\!\zeta({\boldsymbol{ \varphi} })>0,} \\ 
			{\!\!0,}&{\!\!\!\zeta({\boldsymbol{ \varphi} })=0,}
	\end{array}} \right.
\end{equation}
where $Z \!=\!({\mu({\boldsymbol{ \varphi} })-{\hat{R}}^+})/{\zeta({\boldsymbol{ \varphi} })}$, $F(Z) \!=\!\frac{1}{\sqrt{2\pi}}\int_{ - \infty }^Z {{e^{ - \frac{{{t^2}}}{2}}}{\rm d}t}$ denotes the cumulative distribution function, and $f(Z)=\frac{1}{\sqrt{2\pi}}{e^{ - \frac{{{Z^2}}}{2}}}$ denotes the probability density function.
Given the non-continuous constraints for each $\varphi_{mb}$ in  (\ref{Problema}) and (\ref{Problemb}), 
we transform the feasible domain for each $\varphi_{mb}$ into a continuous region, given by \vspace{-1.5 mm}
\begin{equation}\label{domain}\vspace{-1.5 mm}
	\mathcal{R}_{mb} = \Big[ {\frac{{\varphi _{m(b-1)}^+  + {\varphi _{mb}^+}}+\delta}{2} ,\frac{{\varphi _{m(b+1)}^+  + {\varphi _{mb}^+}}-\delta}{2} } \Big],
\end{equation}
where ${\varphi _{mb}^+}$ corresponds to the current best observation point ${\boldsymbol \varphi}^{(i^+)}$, $\varphi_{m0}^+\! =\! -\varphi_{m1}^+$ and $\varphi_{m(B+1)}^+ \!=\! 4\pi-\varphi_{mB}^+$ handle the boundary cases for $b \! =\! 1$ and $b \! = \! B$, respectively.

As a result, the next sample point ${\boldsymbol{ \varphi} }^{(S+1)}$ is selected by \vspace{-1.5 mm}
\begin{equation}\label{newsample} \vspace{-1.5 mm}
	{  {\boldsymbol{ \varphi} }^{(S+1)}}=  \mathop {\arg \max }\limits_{  {\boldsymbol{ \varphi} }\in \prod_{m = 1}^M {\prod_{b = 1}^B {{\mathcal{R}_{mb}}} } } \mathcal{L}({\boldsymbol{ \varphi} }).
\end{equation}
Problem (\ref{newsample}) allows for straightforward computation of the first- and second-order derivatives, which thus can be efficiently solved with continuous optimization methods, such as the quasi-Newton method \cite{9784826}.

The BO-based algorithm for solving problem (\ref{Problem}) is summarized in Algorithm \ref{algorithm1}.
We assume that problem (\ref{newsample}) is solved by using the quasi-Newton method, with a complexity order of $\mathcal{O}(T  BM  \tilde{K}^3 \Upsilon)$, where \(T\) denotes the number of iterations required for convergence and $\tilde{K}={\rm max}(K_1,\cdots,K_\Upsilon)$.
The complexity order of Algorithm \ref{algorithm1} is $\mathcal{O}(L  (T  BM  \tilde{K}^3 \Upsilon + |\mathcal{D}_{S+L}|^3))$, where \(|\mathcal{D}_{S+L}|\) denotes the cardinality of $\mathcal{D}_{S+L}$ and accounts for the complexity of updating the posterior distribution in (\ref{predictive distribution}). 
\begin{algorithm}[!t]
	\renewcommand{\thealgorithm}{1}
	\caption{BO-based algorithm for solving problem (\ref{Problem})}\label{algorithm1}
	\begin{algorithmic}[1]
		\STATE \textbf{Initialization:} Randomly select $S$ collective rotation angle vectors to construct a sample set $\mathcal{D}_S$.
		\STATE {\textbf{for}} $\iota=1:L$ {\textbf{do}} 
		\STATE \hspace{0.1cm} Obtain $\mathcal{R}_{mb},\forall m\!\in\!\mathcal{M},  b\in\mathcal{B}$ via (\ref{domain}).
		\STATE \hspace{0.1cm} Select a new sample ${{\boldsymbol{ \varphi} }^{(S+\iota)}}$ by solving problem (\ref{newsample}).
		\STATE \hspace{0.1cm} Obtain a new observation ${\hat{R}}({{\boldsymbol{ \varphi} }^{(S+\iota)}}) $ via (\ref{sum-rate2}).
		\STATE \hspace{0.1cm} Augment $\mathcal{D}_{S+\iota}=\{\mathcal{D}_{S},({{\boldsymbol{ \varphi} }^{(S+\iota)}},{\hat{R}}({{\boldsymbol{ \varphi} }^{(S+\iota)}}) )\}$.
		\STATE \hspace{0.1cm} Update posterior distribution of $ {\hat{R}}^*({  {\boldsymbol{ \varphi} }^{(S+\iota+1)}}) $ via (\ref{predictive distribution}).
		\STATE {\textbf{end for}}
		\STATE \textbf{return}  ${\boldsymbol \varphi}^{(i^+)}$ with $i^+ \!=\! \arg \max_{i \in \{1, \cdots, S+L\}} {\hat{R}}({\boldsymbol \varphi}^{(i)})$.
	\end{algorithmic}
	\label{alg1}
\end{algorithm}
\vspace{-3 mm}
\section{Simulation Results}
\vspace{-1 mm}
In this section, we evaluate our proposed 6DMA-aided cell-free mMIMO system in terms of the average achievable sum-rate. 
Unless otherwise stated, the simulation settings are as follows.
We set $B\! =\! 6$, $N\! = \!2\! \times\! 1$, $r\! =\! 1\!~\rm m$, $h\! =\! 10\!~\rm m$, $\lambda\!=\!0.125\!~\rm m$, $\nu_0\!=\!-40\!~\rm dB$, $P_0\!=\!1\!~\rm mW$, $\sigma ^2\!=\!-80\!~\rm dBm$, $\delta = \frac{\lambda}{2r}$, $\Upsilon=100$ and $L\!=\!100$ (for Algorithm \ref{algorithm1}).
According to the 3GPP standard \cite{3GPP}, the 3D beam pattern of the directional antenna is given by $g\left( {\varphi _{mb}} \right) = {\bar g\left( {{\varphi _{mb}}} \right)}/{\eta}$, with\vspace{-1.5 mm}
\begin{equation}\vspace{-1.5 mm}
	\bar g\left( {{\varphi _{mb}}} \right)\left| {{\text{dBi}}} \right. = {{{{A_m} \!-\! \min \{ { - \left( {{A_h}({\varphi _{mb}})\! +\! {A_v}} \right),{\text{S}}{{\text{L}}_h}} \}}},}
\end{equation}
where ${A_h}({\varphi _{mb}})\!\left| {{\text{dBi}}} \right.\! =\!  -\! \min \{ {12{{\big( {\frac{{{\phi _{mk}} - {\varphi _{mb}}}}{{{\phi _{{\text{3dB}}}}}}} \!\big)}^2}\!,{\text{S}}{{\text{L}}_h}} \}$ and ${A_v}\left| {{\text{dBi}}} \right. \!=\!  - \min \{ {12{{\big( {\frac{{{\theta _{mk}} - {\theta _{\text{tilt}}}}}{{{\theta _{{\text{3dB}}}}}}} \big)}^2},{\text{S}}{{\text{L}}_v}} \}$ denote the horizontal and vertical radiation patterns, respectively, ${\phi _{{\text{3dB}}}}\!=\!{\theta _{{\text{3dB}}}}\!=\!65^ \circ$ denotes the ${\text{3-dB}}$ beamwidth, ${\text{S}}{{\text{L}}_h}\!=\!{\text{S}}{{\text{L}}_v}\!=\!20\!~{\rm dBi}$ denotes the side lobe level, ${\theta _{\text{tilt}}}\!=\!100^\circ$ denotes the electrical antenna downtilt, $A_m\! =\! 14\!~\rm dBi$ denotes the maximum antenna gain, and $\eta=\frac{1}{{4\pi }}\int_0^{2\pi } {\int_0^\pi  {\bar g\left( {{\varphi _{mb}},{\phi _{mk}},{\theta _{mk}}} \right)\sin ({\theta _{mk}}){\text{d}}{\theta _{mk}}{\text{d}}{\phi _{mk}}} }$ denotes the normalization factor.

\begin{figure}[t]
	\centering
	\includegraphics[width=3.5in]{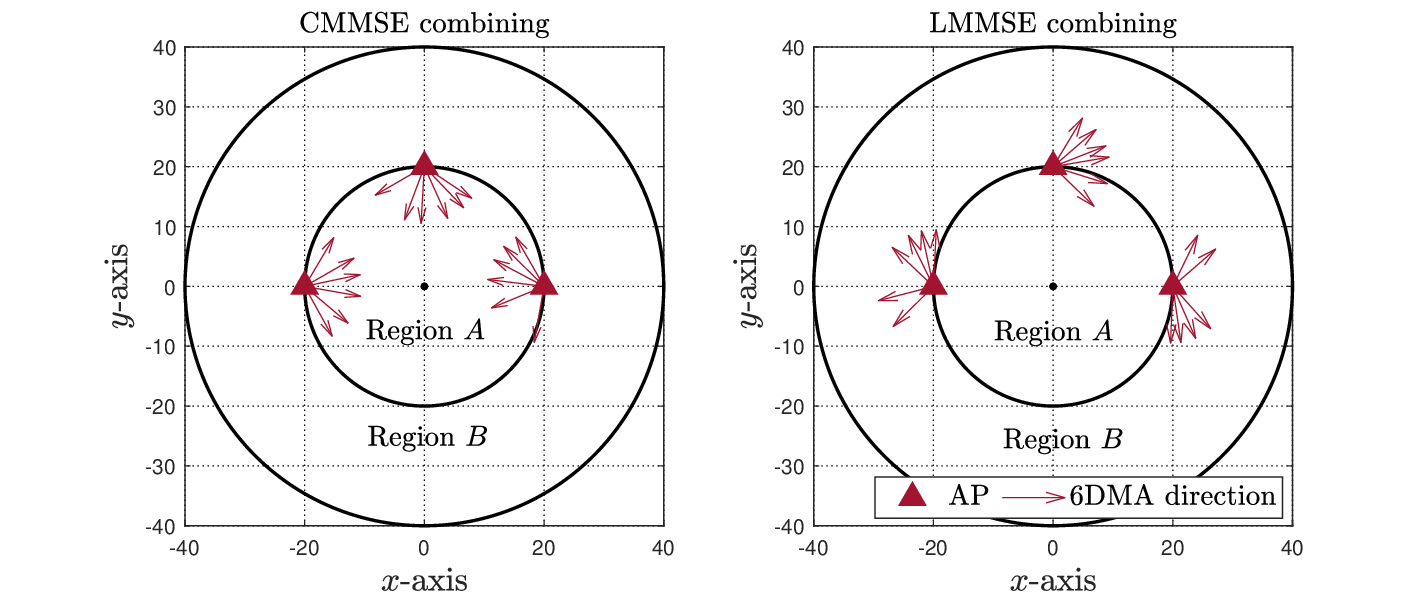}
	\vspace{-7 mm}
	\caption{Optimized rotations of 6DMA surfaces using the CMMSE and LMMSE receiver combining methods.} 
	\vspace{-5 mm}
	\label{result1}
\end{figure}

For the users' spatial distribution $\Phi$, we assume users are i.i.d. over two concentric circular regions, denoted by region A and region B, centered at the global CCS origin, with radii $ R_A \!=\! 20 \, \text{m}$ and $ R_B \!=\! 40 \, \text{m} $, respectively (see Fig. \ref{result1}).
The average number of users, i.e., $\bar{K} = \mu_A\pi R_A^2 + \mu_B \pi (R_B^2-R_A^2)$, is set to 30, where $\mu_A$ and $\mu_B$ denote the user densities in regions A and B, respectively.
We consider $M\! =\! 3$ APs, which are positioned along region A's boundary, with azimuth angles of $0$, $\frac{\pi}{2}$, and ${\pi}$ relative to the positive $x$-axis of the global CCS.

We consider the following benchmark schemes:{\footnote{The total number of antennas in each benchmark scheme is the same as that of the proposed scheme.}}
\begin{itemize}
\item \textbf{Cell-free mMIMO with half-space isotropic 6DMA:}
		This scheme considers a cell-free mMIMO system with $M$ 6DMA-APs, each equipped with half-space isotropic antennas that radiate uniformly within a hemisphere, i.e., $g\left( {{\phi _{mk}},{\varphi _{mb}}} \right) \!=\! 2\cdot{\bf 1}({{\cos}}\left( {{\phi _{mk}} - {\varphi _{mb}}} \right)\! >\! 0)$.
\end{itemize}
\begin{itemize}
	\item \textbf{Centralized mMIMO with directional 6DMA:}  
	This scheme considers a centralized mMIMO system with a single 6DMA-AP located at the  global CCS origin, equipped with $B$ surfaces. Each surface consists of $MN$ directional antennas.
\end{itemize}  
\begin{itemize}
	\item \textbf{Cell-free mMIMO with directional sectorized UPA:}
	This scheme considers a cell-free mMIMO system with $M$ FPA-APs, each equipped with three sectorized UPAs.
	Each UPA consists of  ${\frac{NB}{3}}$ directional antennas.
\end{itemize}
\begin{itemize}
	\item \textbf{Cell-free mMIMO with isotropic ULA}:
	This scheme considers a cell-free mMIMO system with $M$ FPA-APs, each equipped with a horizontally arranged uniform linear array (ULA) consisting of ${{NB}}$ isotropic antennas.
\end{itemize}

\begin{figure}[t]
	\centering
	\includegraphics[width=3.2in]{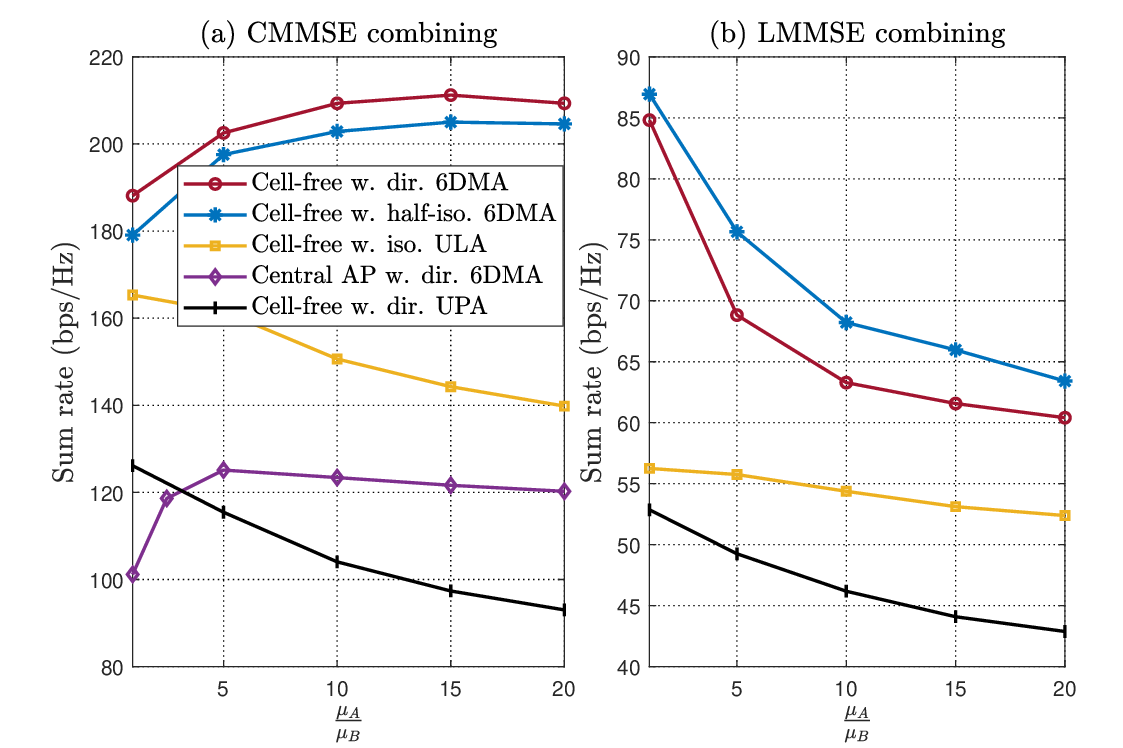}
	\vspace{-4 mm}
	\caption{Average achievable sum-rate versus user density ratio with different receiver combining methods.} 
	\vspace{-3 mm}
	\label{result2}
\end{figure}

\begin{figure}[t]
	\centering
	\includegraphics[width=3.2in]{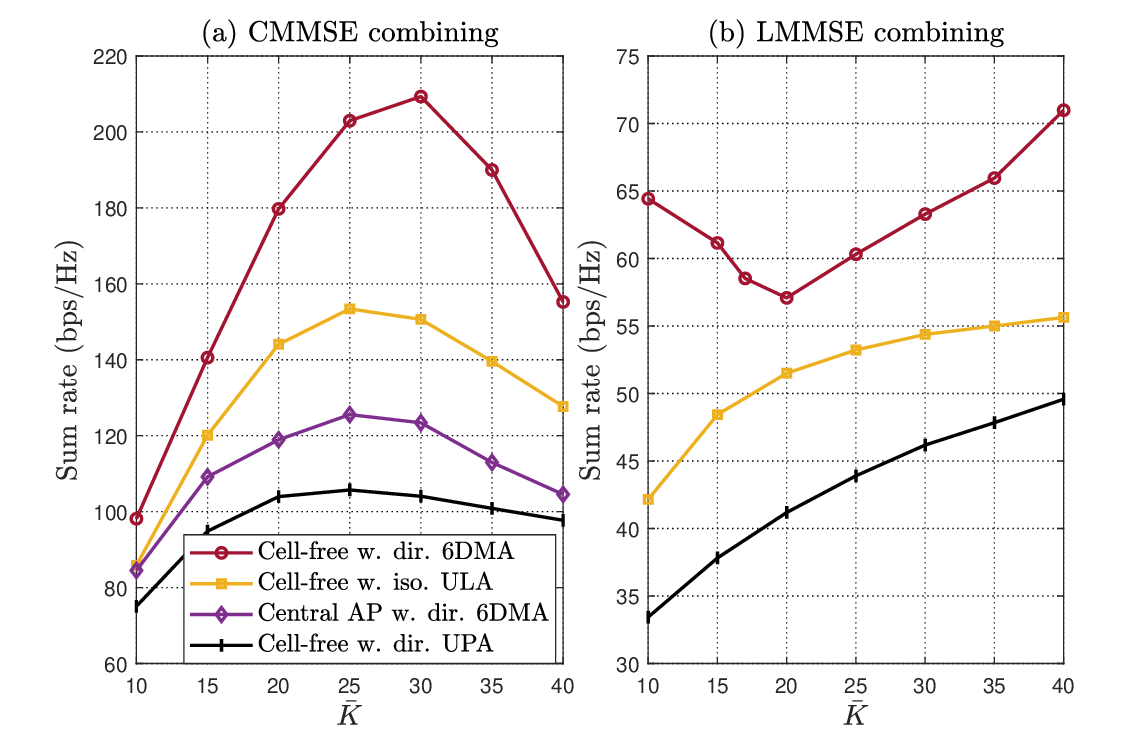}
	\vspace{-4 mm}
	\caption{Average achievable sum-rate versus average number of users with different receiver combining methods.} 
	\label{result3}
\end{figure}
Fig. \ref{result1} shows visual representations of the optimal rotations of 6DMA surfaces using the CMMSE and LMMSE receiver combining methods, with a user density ratio of $\frac{\mu_A}{\mu_B} = 5$. It is observed that with the CMMSE combining, the 6DMA surfaces tend to rotate towards the higher-density user region (i.e., region A). 
In contrast, with the LMMSE combining, the 6DMA surfaces avoid pointing towards region A to reduce the interference among users, due to the lack of global CSI for designing the optimal combining vectors at all APs. 
The above results show the necessity of flexible rotation of 6DMA surfaces using different receiver combining methods in the cell-free mMIMO systems.

Fig. \ref{result2} shows the average achievable sum-rate versus the user density ratio $\mu_A/\mu_B$ with different receiver combining methods. 
It is observed that the proposed 6DMA-aided cell-free mMIMO schemes with both directional and half-space isotropic antennas outperform the other benchmark schemes. 
Note that although the centralized 6DMA-AP can also rotate its surfaces based on the users' spatial distribution, it cannot leverage the macro spatial diversity provided by the cell-free mMIMO system, and even results in worse performance than the ULA-aided cell-free mMIMO system.
In addition, with the LMMSE combining, the half-space isotropic 6DMA outperforms its directional counterpart, due to its broader angular coverage, which reduces channel correlation among different users  and enhances local CSI-based interference suppression.

Fig. \ref{result3} illustrates the average achievable sum-rate versus the average total number of users, $\bar K$, with different receiver combining methods.
In the case of CMMSE combining, the sum-rate initially increases with $\bar K$ due to the higher spatial multiplexing gain achieved by the 6DMA surfaces at all APs. 
However, when $\bar K$ exceeds the total number of antennas, the sum-rate begins to decline due to more severe multiuser interference.
In contrast, in the case of LMMSE combining, each AP independently performs MMSE combining based on its local CSI.
As a result, when $\bar{K}$ is small and the users are sparsely distributed, the local CSI-based receiver combining is ineffective to suppress the multiuser interference, thus leading to degraded sum-rate.
As $\bar{K}$ increases, the multi-user diversity effect becomes more significant, which renders some users having low channel correlation with the other users and thus helps improve the users' sum-rate.

\vspace{-2 mm}
\section{Conclusions}
\vspace{-1 mm}
In this letter, we considered a new 6DMA-aided cell-free mMIMO system where distributed APs are equipped with 6DMA surfaces which can be flexibly moved along a circular track subject to minimum-angle constraints.
To maximize the average achievable sum-rate of users, we proposed a BO-based algorithm to jointly optimize the rotation angles of all 6DMA surfaces at all APs based on the users' spatial distribution.
Numerical results verified that the proposed system can achieve superior performance compared to the centralized mMIMO system with 6DMAs and the FPA-based cell-free mMIMO system, especially when the users' spatial distribution is more non-uniform.
In addition, it was revealed that the optimized 6DMA rotations at APs vary significantly using global versus local CSI-based receiver combining methods.
\vspace{-3 mm}
\bibliographystyle{IEEEtran}
\bibliography{123}
\end{document}